\documentclass[aps,prl,twocolumn,a4paper,showpacs]{revtex4}
\usepackage{amsmath}
\usepackage{amssymb}
\usepackage{amsfonts}
\usepackage{times,txfonts}
\usepackage{color}
\usepackage{bm}
\usepackage{graphicx}

\newcommand{\bld}[1]{\textbf{#1}}
\newcommand{\red}[1]{\textcolor{red}{#1}}
\newcommand{\green}[1]{\textcolor{green}{#1}}
\newcommand{\blue}[1]{\textcolor{blue}{#1}}
\newcommand{\magnta}[1]{\textcolor{magenta}{#1}}

\begin{document}

\title{Rotational Response of Two-Component Bose-Einstein Condensates in Ring Traps}

\author{P. L. Halkyard}
\affiliation{Department of Physics, Durham University, Rochester Building, South Road, Durham DH1 3LE, United Kingdom}

\author{M. P. A. Jones}
\affiliation{Department of Physics, Durham University, Rochester Building, South Road, Durham DH1 3LE, United Kingdom}

\author{S. A. Gardiner}
\affiliation{Department of Physics, Durham University, Rochester Building, South Road, Durham DH1 3LE, United Kingdom}


\date{\today}
\begin{abstract}
We consider a two-component Bose-Einstein condensate (BEC) in a ring trap in a rotating frame, and show how to determine the response of such a configuration to being in a rotating frame, via accumulation of a Sagnac phase.  This may be accomplished either through population oscillations, or the motion of spatial density fringes. We explicitly include the effect of interactions via a mean-field description, and study the fidelity of the dynamics relative to an ideal configuration.
\end{abstract}
\maketitle

\begin{figure}[t]
\centering
\includegraphics[width=8.8cm]{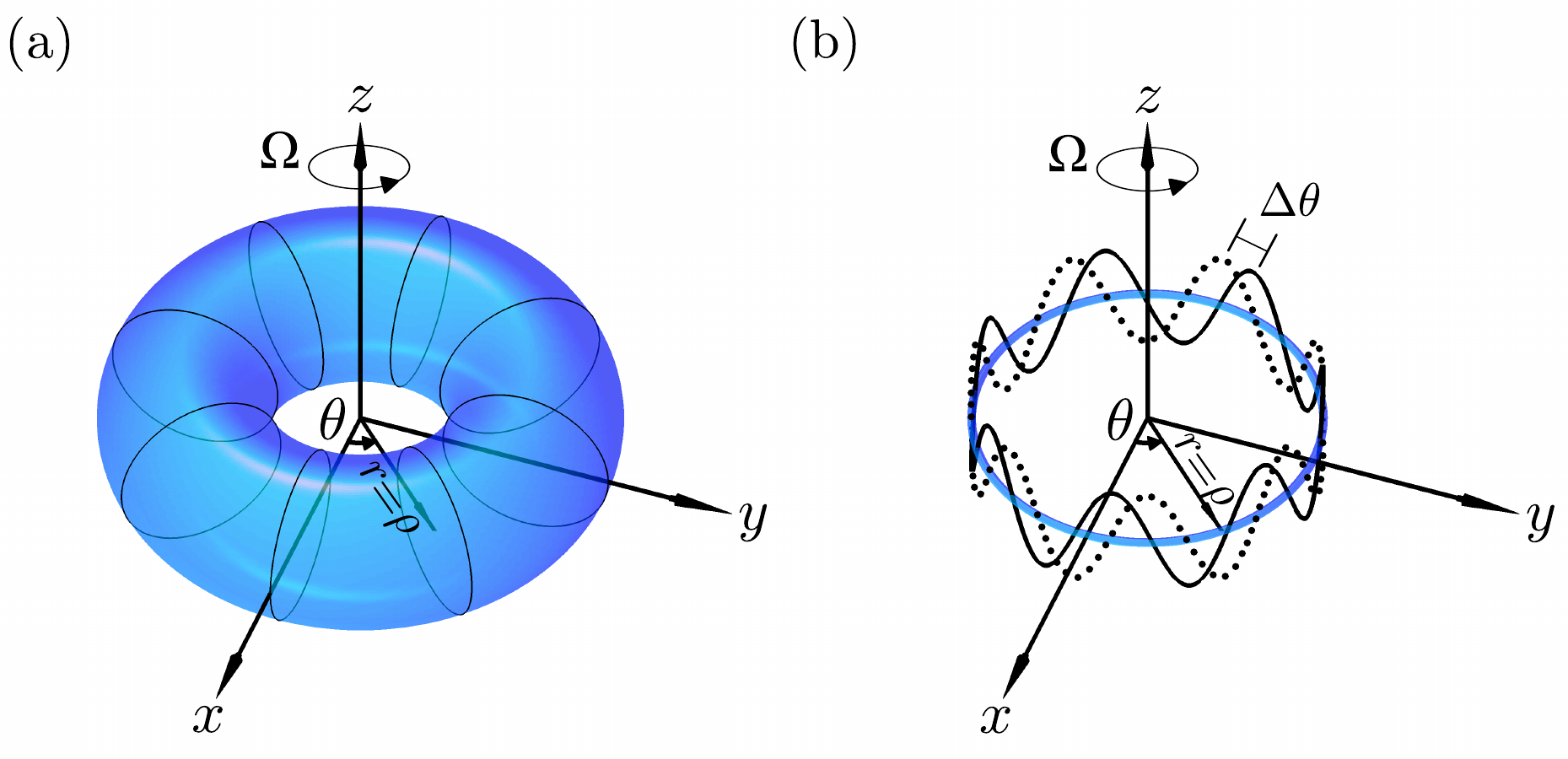}
\caption{(color online). (a) Equipotential surface of an axially symmetric toroidal trapping potential with radius $\rho$, in a frame rotating with angular frequency $\Omega$ about the axis of symmetry ($z$ axis).  (b) In the quasi-1D limit, the equipotential surface shrinks to a ring, and the trapped atoms' spatial dynamics are in terms of the angle $\theta$ only.  
Spatial fringes in the macroscopic atomic wavefunction will shift by $\Delta\theta = \Omega$ multiplied by the interrogation time.} 
\label{fig:torus}
\end{figure} 
The Sagnac effect 
\cite{Post1967} 
is a rotational phenomemon describing the phase shift, $\Delta\theta$, between two coherent, counter-propagating waves traversing the same, closed path in a rotating frame.  Originally discovered as an optical effect, it is actually more universal \cite{Varoquaux2008}; it has been observed in matter-wave interferometry experiments aiming to make precision measurements of rotation \cite{Wu2007}, and has even been proposed as a method of testing general relativity \cite{Dimopoulos2008}.  

An attractive and theoretically simple geometry for observing Sagnac-like effects in matter-waves
is a ring trap, and with the excellent degree of control and precision now available over magnetic and laser fields, the creation of such traps has recently been achieved by a number of groups worldwide \cite{Ryu2007,Henderson2009,Arnold2006,Sauer2001}.  Recent experiments with Bose-Einstein condensates (BECs) in ring traps \cite{Ryu2007} show how the coherent transfer of orbital angular momentum to a trapped BEC \cite{Andersen2006,Allen1992} can induce long-lived, superfluid flow.  Two different flows (usually considered to be counter-propagating, although this is not strictly necessary --- as we will show, one of the flows may for example be zero) are required to observe the Sagnac effect in an atom-optical context.  We show that there are a number of advantages in using a two-component BEC \cite{Pu1998}, made up of a single atomic species with two relevant internal states, particularly in ameliorating the frequently problematic effects of atom-atom interactions.  It is therefore not necessary to assume negligible mean-field interactions \cite{Thanvanthri2009} to cleanly observe the rotational response brought out by our proposed protocols.
We describe how the accumulation of a Sagnac phase can then be observed both through population oscillations between the internal states, and by precession of density fringes within a particular internal state.  
We show how, in the case of density fringes, mean-field interactions from one component can stabilize the fringes in the other if the scattering lengths are approximately equal (as in, e.g., $^{87}$Rb \cite{Boxio2008}). Hence, within a mean-field picture, the repulsive interactions can in principle be arbitrarily strong without affecting the interferometric signal.  Finally, we discuss the sensitivity of our protocols. 
  
We consider a 2-component BEC composed of a single species with two relevant internal states,  confined within an axially symmetric toroidal trapping potential considered to be insensitive to the internal state [Fig.\ \ref{fig:torus}(a)].  We employ a mean-field treatment, describing the sample with two coupled Gross-Pitaevskii equations:
\begin{equation} \label{cgpe}
i\hbar\frac{\partial\Psi_j}{\partial t} =
\Biggl[-\frac{\hbar^2\nabla^2}{2M} + V + (-1)^j\frac{\hbar\omega}{2}
+ 
\frac{4\pi \hbar^{2} N}{M}\sum_{k=1}^{2}a_{jk} |\Psi_k|^2\Biggr]\Psi_j,  
\end{equation} 
where $\Psi_j(\bld{r})$ is the macroscopic wavefunction for atoms in internal state $j$ [normalised such that $\sum_{j=1}^{2}\int d^3\bld{r}|\Psi_j(\bld{r})|^2=1$], $N$ is the total atom number, $\hbar\omega$ is the energy difference between the two internal states, the $a_{jk}$ are the $s$-wave scattering lengths, and $M$ is the atomic mass. In terms of cylindrical coordinates $\{r,\theta,z\}$, we take the confining potential to be $V(\bld{r}) = M[ \omega_r^2(r-\rho)^2+\omega_z^2z^2]/2$, with torus radius $\rho$ and angular trapping frequencies $\omega_{r}$, $\omega_{z}$.  Assuming sufficiently tight radial and axial confinement the dynamics in these directions are ``frozen out,'' permitting a quasi-1D description \cite{foot1,Bagnato1991}
[see Fig.\ \ref{fig:torus}(b)].  Projecting out the $\{r,z\}$ dependences and moving to a frame rotating counterclockwise about the $z$-axis with angular frequency $\Omega$ transforms Eq.\ (\ref{cgpe}) to 
\begin{equation} \label{cgpe1ddim}
i\frac{\partial\psi_j}{\partial t} =  \left[-\frac{1}{2}\frac{\partial^2}{\partial \theta^2} - i\Omega\frac{\partial}{\partial \theta} + (-1)^{j}\frac{\omega}{2}
+\sum_{k=1}^{2}g_{jk}|\psi_k|^2\right]\psi_j,  
\end{equation}
where the time is now in units of $\tau = M\rho^{2}/\hbar$, and frequencies are in units of $\tau^{-1}$.
The $g_{jk}=2MN\rho a_{jk}\sqrt{\omega_r\omega_z}/\hbar$, and the normalization condition is $\sum_{j=1}^{2}\int_0^{2\pi} d\theta|\psi_{j}(\theta)|^2=1$.  

\begin{figure}[t]
\centering
\includegraphics[width=8.8cm]{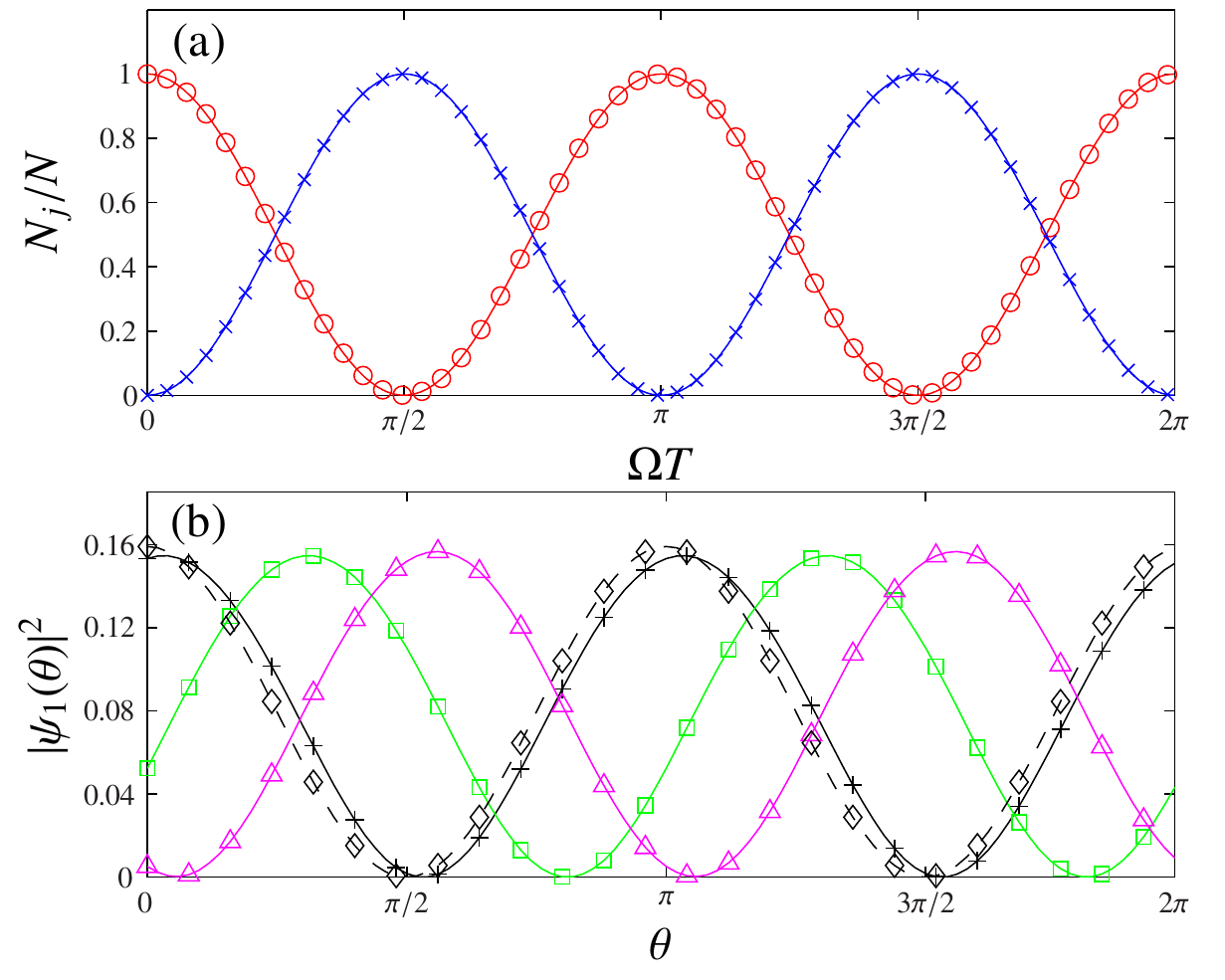}
\caption{(color online). (a) State $j$ mean relative population $N_j/N$ as a function of interrogation time $T$, for  (\red{$\circ$}) $j=1$ and (\blue{$\times$}) $j=2$,  following 
an $S_{N}(T)$ sequence, 
with $\ell=1$.  (b) Angular density fringes of state $1$, with $\ell=1$, $m=0$, $\Omega=1$, for  ($\diamond$)  $T = 0$, ($+$)  $T = 0.1$,  (\green{$\square$})  $T = 1$, and  (\magnta{$\bigtriangleup$})  $T = 8.3$, following an $S_{F}(T)$ sequence.   The interaction strengths $g_{jk}$ correspond to an $^{87}$Rb configuration. All quantities are dimensionless.}
\label{fig:signal}
\end{figure}

To describe a BEC composed of atoms in a coherent internal superposition state,  we introduce the vector notation
\begin{equation}
\vec{\psi}(\theta) =
\left(
\begin{array}{c} 
\psi_{1}(\theta) \\ \psi_{2}(\theta) 
\end{array}
\right).
\end{equation}
We now present our Sagnac interferometry protocols.  We assume all atoms to be initially in internal state $1$ and in the motional ground state, such that the initial state $\vec{\psi}^{I}$ has $\psi_{1}^{I}=1/\sqrt{2\pi}$, $\psi_{2}^{I} = 0$. Applying a resonant $\pi/2$ pulse (a ``splitting'' pulse) to the internal two-state transition then yields
$\vec{\psi}^{\pi/2}(\theta) = U_{\pi/2}\vec{\psi}^{I}$, where
\begin{equation}
\vec{\psi}^{\pi/2}(\theta) = \frac{1}{2\sqrt{\pi}}
\left(
\begin{array}{c} 
1 \\ 
1
\end{array}
\right),
\quad
U_{\pi/2} =
\frac{1}{\sqrt{2}}
\left(
\begin{array}{rr} 
1 & 1 \\ 
1 & -1 
\end{array}
\right).
\end{equation}
We imprint different angular momenta onto the spatial modes associated with the two internal states (e.g., by transfer of orbital angular momentum of light \cite{Andersen2006}), producing
$\vec{\psi}^{\ell m}(\theta) = U_{\ell m}\vec{\psi}^{\pi/2}(\theta)$, where
\begin{equation}
\vec{\psi}^{\ell m}(\theta) = \frac{e^{im\theta}}{2\sqrt{\pi}}
\left(
\begin{array}{c} 
e^{i\ell\theta} \\ 
e^{-i\ell\theta}
\end{array}
\right),
\quad
U_{\ell m} =
\left(
\begin{array}{ll} 
e^{i(m+\ell)\theta} & 0 \\ 
0 & e^{i(m-\ell)\theta} 
\end{array}
\right).
\end{equation}
With this notation we can describe the symmetric case ($m=0$), the case where angular momentum is imprinted on $\psi_{1}$ only ($m=\ell$), and also permissable intermediate cases ($\{m,\ell\}$  both either integer or half-integer).  A free evolution $f(T/2)$ follows 
[$f(t)$ denotes an evolution governed by Eq.\ (\ref{cgpe1ddim}) for a time  $t$, and $T/2$ is half the total interrogation time].  As the atom fields have uniform density, this takes a very simple form:
\begin{equation}
\vec{\psi}^{T/2}(\theta)
=
\frac{e^{-i\varphi_{1}T/2}e^{im(\theta+\Omega T/2)}}{2\sqrt{\pi}}
\left(
\begin{array}{c} 
e^{i\varphi_{2}T/2}e^{i\ell(\theta+\Omega T/2)} \\ 
e^{-i\varphi_{2}T/2}e^{-i\ell(\theta+\Omega T/2)}
\end{array}
\right),
\end{equation}
where 
$
\varphi_{1} = (m^{2}+\ell^{2})/2 +(g_{11}+2g_{12}+g_{22})/8\pi
$,
and
$
\varphi_{2} = \omega/2 - 2m\ell +(g_{22}-g_{11})/8\pi
$.
We apply a $\pi$ pulse  
$U_{\pi}$, 
which swaps the two components, and allow another free evolution $f(T/2)$ (completing the total interrogation time $T$).  This negates the accumulated relative phase described by $\varphi_{2}$, producing
\begin{equation}
\vec{\psi}^{T}(\theta)
=
\frac{e^{-i\varphi_{1}T}e^{im(\theta+\Omega T)}}{2\sqrt{\pi}}
\left(
\begin{array}{c} 
e^{-i\ell(\theta+\Omega T)} \\ 
e^{i\ell(\theta+\Omega T)}
\end{array}
\right).
\label{Eq:psiT}
\end{equation}
We repeat the angular-momentum-imprinting procedure $U_{\ell m}$, which, due to the application of $U_{\pi}$ at $t=T/2$, undoes the relative difference in angular momentum between the spatial modes associated with the two internal states.  
Following this by a second $\pi/2$ pulse (a ``recombination'' pulse) produces 
$\vec{\psi}^{R}(\theta) = U_{\pi/2}U_{\ell m}\vec{\psi}^{T}(\theta)$, 
where
\begin{equation}
\vec{\psi}^{R}(\theta)
=
\frac{e^{-i \varphi_{1} T}e^{im(2\theta+\Omega T)}}{\sqrt{2\pi}}
\left(
\begin{array}{cc}
\cos(\ell\Omega T)
\\
-i\sin(\ell\Omega T)
\end{array}
\right),
\end{equation}
and there is no subsequent change to the populations.  We summarize this sequence by $S_{N}(T) \equiv U_{\pi/2}U_{\ell m}f(T/2)U_{\pi}f(T/2)U_{\ell m}U_{\pi/2}$.  The value of $\Omega$ is inferred from the population, e.g., in internal state 2, $N_{2}=N[1 - \cos(2\ell\Omega T)]/2$; the populations oscillate with period $\pi/\ell\Omega$  [see Fig.\ \ref{fig:signal}(a)]. Conveniently, any experimentally significant change of $N_{2}$ from zero is a clear signal for finite $\Omega$.  Note that the same response is obtained when angular momentum is imprinted on $\psi_{1}$ only ($m=\ell$) as for the symmetric case ($m=0$).  It is thus not essential to imprint angular momentum on both components, permitting a significant experimental simplification.

\begin{figure}[t]
\centering
\includegraphics[width=8.4cm]{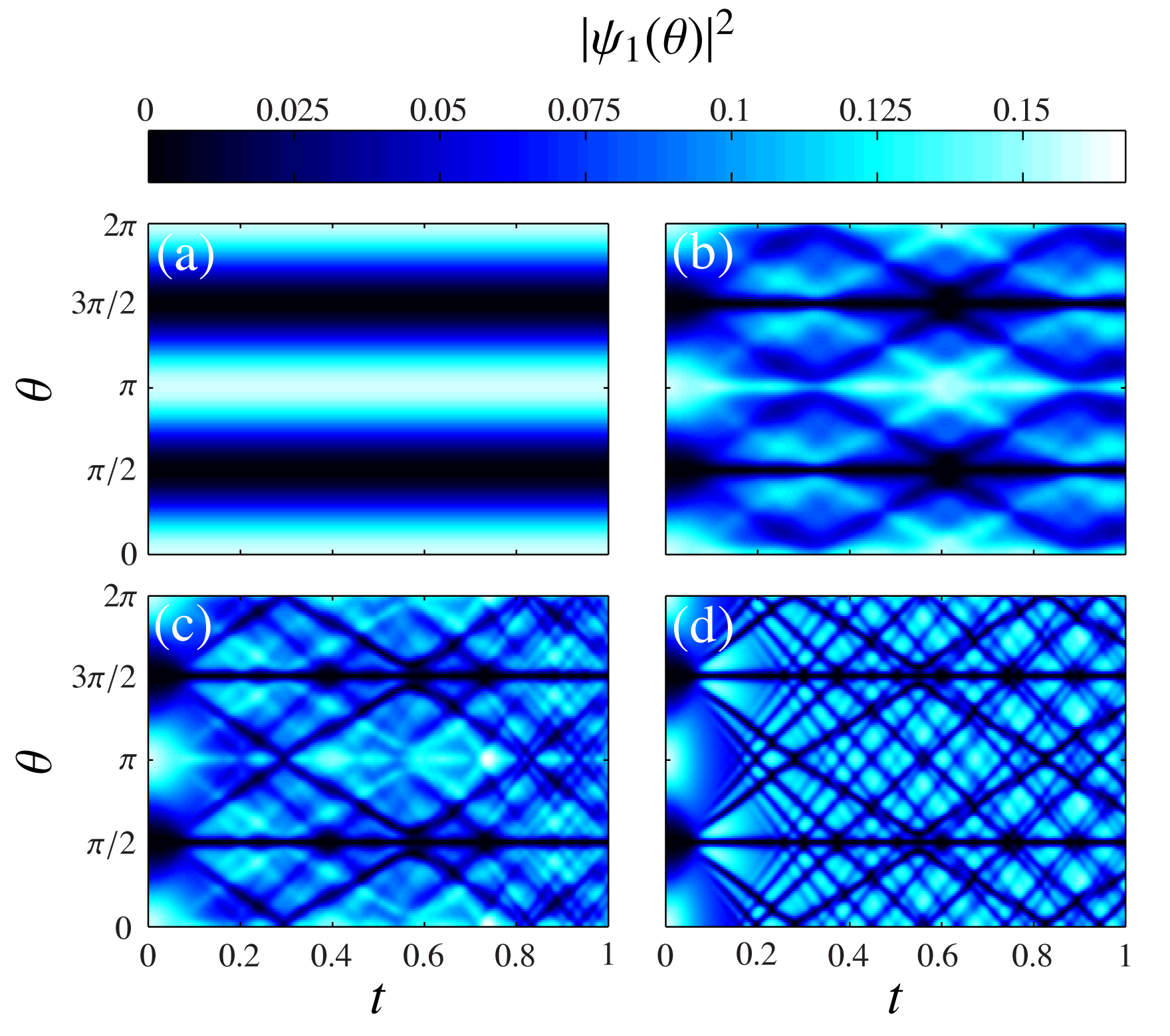}
\caption{(color online). Component 1 angular position density $\left|\psi_1(\theta,t)\right|^2$ evolved by Eq.\ (\ref{cgpe1ddim}) from Eq. (\ref{Eq:InitialSine}) for $m=0$, $\ell=1$ $\Omega=0$, $g_{11}=g_{22}=1000$, with: (a) $g_{12} = 1000$, (b) $g_{12}=750$, (c) $g_{12}=500$, (d) $g_{12}=0$.  Note that $\left|\psi_2(\theta)\right|^2$ is identical except for a $\pi$ shift in $\theta$, and finite $\Omega$ causes the densities to precess by $\Delta\theta =\Omega t$. All quantities are dimensionless.}
\label{fig:RT}
\end{figure}

Alternatively, we may omit the second application of $U_{\ell m}$, instead applying $U_{\pi/2}$ directly to $\vec{\psi}^{T}(\theta)$ to produce 
\begin{equation}
\vec{\psi}^{R^{\prime}}(\theta)
=
\frac{e^{-i \varphi_{1}T}e^{im(\theta+\Omega T)}}{\sqrt{2\pi}}
\left(
\begin{array}{cc}
\cos(\ell[\theta+\Omega T])
\\
-i\sin(\ell[\theta+\Omega T])
\end{array}
\right).
\label{Eq:AlternativeInterferometer}
\end{equation}
The atomic angular density in, e.g., internal state $1$ is therefore 
$N|\psi_{1}(\theta)|^{2}=N\{1 + \cos(2\ell[\theta + \Omega T])\}/4\pi$; 
the fringe spacing is $\pi/\ell$, and the fringe positions change with the total interrogation time $T$ with rate $\Omega$.  Hence, measurable information about $\Omega$ can be obtained without repeated angular momentum imprinting.  If the measurement is not immediate, however, the $\psi_{j}(\theta)$ continue to evolve.  Simplifying to the case where $g_{11}=g_{12}=g_{22}=g$, for an initial condition 
\begin{equation}
\psi_{1}(\theta) = \frac{e^{im\theta}\cos(\ell\theta)}{\sqrt{2\pi}},
\quad 
\psi_{2}(\theta) = -\frac{ie^{im\theta}\sin(\ell\theta)}{\sqrt{2\pi}} 
\label{Eq:InitialSine}
\end{equation}
[e.g., formed from $\vec{\psi}^{I}$ by a $U_{\pi/2}U_{\ell m}U_{\pi/2}$ sequence, or equivalent to Eq.\ (\ref{Eq:AlternativeInterferometer}) with $\theta$ redefined and the global phase $e^{-i\varphi_{1}T}$ discarded] the mean-field contributions to Eq.\ (\ref{cgpe1ddim}) are 
$
g[\cos^{2}(\ell\theta) + \sin^{2}(\ell\theta)]/2\pi \equiv g/2\pi
$.  A subsequent evolution $f(t)$ yields
\begin{equation}
\vec{\psi}^{t}(\theta)
=
\frac{e^{-i \varphi_{1}t}e^{im(\theta + \Omega t)}}{\sqrt{2\pi}}
\left(
\begin{array}{cc}
e^{i\varphi_{2}t}\cos(\ell[\theta+(\Omega-m)t])
\\
-ie^{-i\varphi_{2}t}\sin(\ell[\theta+(\Omega-m)t])
\end{array}
\right),
\label{Eq:SuperpositionEvolution}
\end{equation}
where the phases simplify to $\varphi_{1}=(m^{2}+\ell^{2})/2 + g/2\pi$, $\varphi_{2} = \omega/2-2m\ell$.
Hence, if the $g_{jk}$ are equal, the fringes in the two components stabilize each other, simply precessing around the ring with rate $\Omega-m$.  Note also that the density fringes yielded by an $S_{F}(T)\equiv f(T)U_{\pi/2}U_{\ell m}U_{\pi/2}$ sequence are identical to those from a $U_{\pi/2}f(T/2)U_{\pi}f(T/2)U_{\ell m}U_{\pi/2}$ sequence when $m=0$ \cite{foot5} [see Fig.\ \ref{fig:signal}(b)].

\begin{figure}[t]
\centering
\includegraphics[width=8.4cm]{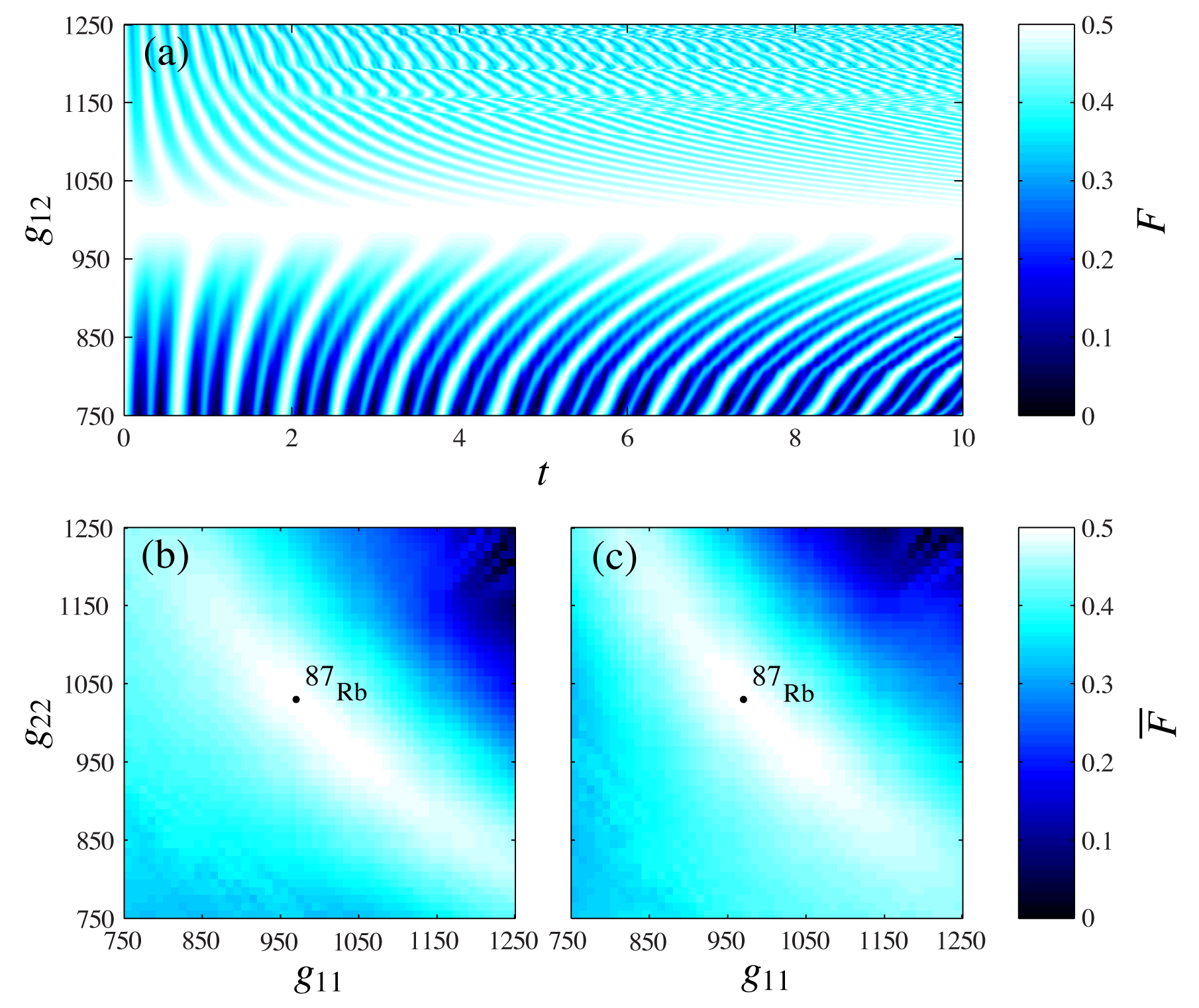}
\caption{(color online). (a) Fidelity $F(t)=|\int_0^{2\pi}d\theta\phi^*(\theta,t)\psi_{1}(\theta,t)|^2$ of $\psi_{1}(\theta,t)$ evolved by Eq.\ (\ref{cgpe1ddim}) from Eq.\ (\ref{Eq:InitialSine}) with $m=0$, $\ell=1$,
$g_{11}=g_{22}=1000$, for varying $g_{12}$	[$\phi(\theta,t)$ is equivalent, but with $g_{11}=g_{12}=g_{22}$].  (b), (c) Time-averaged fidelities $\overline{F}(T)=T^{-1}\int_{0}^{T}dt F(t)$, 
when $g_{12}=1000$, $T=10$, for $j=1,2$; $^{87}$Rb lies within the (white) high-fidelity region.  All quantities are dimensionless, and the maximum possible $F(t)$ is $0.5$.}
\label{fig:stability}
\end{figure}

If the $g_{jk}$ are not equal, the fringe pattern can be strongly disrupted, as shown in Fig.\ \ref{fig:RT}. The scattering lengths can be very similar, however; in the case of $^{87}$Rb, if we consider $F=2$, $m_{F}=1$  to be internal state 1, and $F=1$, $m_{F}=-1$ to be internal state 2, then, in Bohr radii, $a_{11}=95.47$, $a_{12}=98.09$, and $a_{22}=100.44$ \cite{Boxio2008}.  Numerical results for an $S_{F}(T)$ sequence for various $T$, with $g_{11}=974$, $g_{12}=1000$, $g_{22}=1024$, $\ell=1$, $m=0$, are shown in Fig.\ \ref{fig:signal}(b). The $g_{jk}$ correspond to an $^{87}$Rb configuration with, e.g., $\rho=10^{-5}$ m, $\omega_{r}=\omega_{z}=2\pi\times400$ Hz, and $N=2800$, which is consistent with current experimental capabilities \cite{Ryu2007}, and for which the fringe profiles are only slightly perturbed \cite{foot2}.  

Temporarily restricting ourselves to $\Omega=0$, we note that Eq.\ (\ref{cgpe1ddim}) preserves initial periodic symmetry [$\psi_{j}(\theta)=\psi_{j}(\theta + 2\pi/\ell)$], and reflection symmetry [e.g., $\psi_{j}(\theta)=(-1)^{j+1}\psi_{j}(-\theta)$], and that, if $\psi_{j}(\theta,t)$ form a solution to Eq.\ (\ref{cgpe1ddim}), then $e^{-im^{2}t/2}e^{im\theta}\psi_{j}(\theta - mt,t)$ also form a solution.  Furthermore, if $\psi_{j}(\theta,t)$ form a solution to Eq.\ (\ref{cgpe1ddim}) when $\Omega=0$, then $\psi_{j}(\theta+\Omega t,t)$ form a solution to Eq.\ (\ref{cgpe1ddim}) when $\Omega \neq 0$.  Hence, so long as the fringes remain resolvable [i.e., do not break up, as, e.g., in Fig. \ref{fig:RT}(d)], the fringe peaks precess with rate $\Omega-m$, and their form is independent of both $\Omega$ and $m$.  We may therefore set $\Omega=0$, $m=0$ when considering the degree of fringe stabilization for unequal $g_{jk}$ numerically.  In Fig.\ \ref{fig:stability}, for $\ell=1$, we see that there is a broad region in parameter space exhibiting substantial stabilization, and Fig.\ \ref{fig:stability}(b) and Fig.\ \ref{fig:stability}(c) show the $^{87}$Rb parameter regime to be comfortably contained within this region. This result also indicates robustness to a comparable difference in the local particle densities of the two internal states due to non-identical trapping potentials.  Different stability regimes, and the rich dynamics shown in Fig.\ \ref{fig:RT}, could also be explored experimentally with the aid of a suitable Feshbach resonance \cite{Chin2008}.  

As each atom is in a superposition state of the two internal states,  measuring the spatial distribution of atoms in, e.g., state $1$ will on average project $N/2$ atoms into that internal state, with variance $N/4$.  Hence, the standard deviation relative to $N$ is $1/2\sqrt{N}$, which can be considered negligible to the Gross-Pitaevskii level of approximation. An ideal \textit{in situ\/} density measurement \cite{Henderson2009} specific to one internal state destroys the coherence between the internal states, but will in principle not affect the classical fields describing the positional states of the two components.  The time-evolutions of the two components are still governed by Eq.\ (\ref{cgpe1ddim}), and so, evolving Eq.\ (\ref{Eq:InitialSine}), $\psi_{1}(\theta,t)$, $\psi_{2}(\theta,t)$ continue to evolve according to Eq.\ (\ref{Eq:SuperpositionEvolution}) for the ideal $g_{11}=g_{12}=g_{22}$ case \cite{foot4}.  Consequently, assuming ideal, nondestructive measurements and $g_{11}\approx g_{12} \approx g_{22}$, the dynamics due to $\Omega\neq 0$ may be tracked through repeated measurements within the same experimental run.  Specifically, the fringe positions can be ``zeroed'' with a first  measurement, and any subsequent precession monitored by later measurements.  Finally, we note that, although we have assumed perfect axial symmetry throughout, we do not expect the effect of any potential asymmetries or corrugations to be significant if their scale is smaller than that set by the condensate chemical potential \cite{Ryu2007}.

For an optical Sagnac interferometer, the fringe shift relative to the fringe width  $\delta_{L}=4A\Omega/\lambda_{L} c$ is commonly taken as a measure of the rotational sensitivity, where $A$ is the interferometer's enclosed area, $\lambda_{L}$ is the optical wavelength, and $c$ is the speed of light \cite{Post1967}.  The relevant angular shift of the spatial fringes we consider is $\Delta\theta = \Omega T$, and the fringe widths scale as $w=\pi/2\ell$; hence $\delta \equiv\Delta\theta/w = 2 \ell \Omega T/\pi$.  In the population-based protocol, $\delta$ gives the number of instances the population alternates between 0 and $N$ over a range  $[0,T]$ of interrogation times.  It is instructive to now consider a more typical simple Mach-Zehnder (MZ) configuration \cite{Eckert2006}.  The de Broglie relations then yield for atoms of momentum $p$ a wavelength $\lambda=h/p$ and velocity $v=p/M$, and hence $\delta_{\textrm{MZ}}=4A\Omega M/h$.  Sensitivity of response therefore appears entirely determined by the enclosed area $A$, as opposed to the interrogation time $T$ of our protocols.  The time taken between the wavepacket splitting and its recombination in fact sets a natural timescale; in a ring geometry, this is given by $T_{1}= \pi\rho/v$, which, using $v=\hbar\ell/\rho M$ and $A=\pi\rho^{2}$, becomes $T_{1}=(AM/h)(2\pi/\ell)$.  Hence, we may rephrase $\delta_{\textrm{MZ}} = 2 \ell \Omega T_{1}/\pi$, which differs from $\delta$ principally in that the time $T_{1}$ at which it is possible to extract useful information is fully determined by $A$ and $p$, rather than being a free parameter \cite{foot3}.  In an MZ configuration \cite{Eckert2006} we expect $v$ to be greater than that corresponding to small $\ell$ in, e.g., a $\rho=10^{-5}$ m toroidal trap \cite{Ryu2007}, implying moderate $T_{1}$ for relatively large $\rho$.  Our proposal is advantageous in using an intense, monochromatic source, where the usual associated issues of interatomic interactions have been circumvented, and where potentially useful information may be extracted at any time $T$.  Large values of $\ell$ \cite{Thanvanthri2009}, although more challenging to generate, will also enhance the rotational response.   Large $\rho$ and $N$ will aid in imaging; we note that ring traps of radius $\rho = 4.8$ cm \cite{Arnold2006} and $^{87}$Rb condensates with $N\sim 10^{6}$ are achievable. The atomic shot noise also places a fundamental limit on the precision with which the spatial fringes or population oscillations can be measured, and hence $\Omega$ inferred.

In conclusion, we have considered the rotational dynamics of a single-species BEC, with two relevant internal states, within a ring trap configuration, and in a rotating frame.  We have proposed a Sagnac-like interferometric protocol where the rotational sensing is manifest as time-dependent population oscillations, in a way that is insensitive to the atom-atom interactions arising within a mean-field picture.  Simpler protocols involve observing the precession of density fringes around the ring.  The fringes are robust for approximately equal interaction strengths (e.g., $^{87}$Rb), and a range of striking dynamics may also be observed by tuning the interactions with Feshbach resonances.  All of these phenomena are observable within the range of recent experimental advances.

We thank K. Helmerson, W. D. Phillips, S. L. Cornish, A. Gauguet, and I. G. Hughes for stimulating discussions, and the UK EPSRC (Grant No. EP/G056781/1) and Durham University (PLH) for support.


\begin{thebibliography}{99}
\bibitem{foot1}
The $\Psi_j(\bld{r},t)$ approximately factorize as $R(r)\psi_j(\theta,t)\phi(z)$, where $\phi(z)=(m\omega_z/\pi\hbar)^{1/4}\exp(-m\omega_{z}z^{2}/2\hbar)$, $R(r)=(m\omega_r/\pi\hbar r^{2})^{1/4}\exp(-m\omega_r[r-\rho]^{2}/2\hbar)$.  This assumes $\sqrt{\hbar/M\omega_r}\ll\rho$, and $|\ell|\ll\rho/\sqrt{\hbar/M\omega_r}$, where $\hbar \ell$ quantifies the maximum angular momentum (the latter optional simplification ensures the effective radial trapping potential, including a centrifugal term, can be considered identical for angular momenta up to $\pm\hbar\ell$).
\bibitem{foot5}
Where $m$ is otherwise a known integer or half-integer value.
\bibitem{foot2}
This exact configuration is not strongly quasi-1D;  it provides a useful basis for comparison, however, and we do not expect a quasi-1D mean field to be critical to the proposal's success.
\bibitem{foot4} 
The vector notation is misleading, however, as the relative phase information between components is lost.
\bibitem{foot3}
Also $\ell = \rho p/\hbar$ need not be integer valued.
\end{thebibliography}
\end{document}